\input harvmac.tex


\def\unlockat{\catcode`\@=11}

\def\lockat{\catcode`\@=12}

\unlockat


\def\newsec#1{\global\advance\secno by1\message{(\the\secno. #1)}
\global\subsecno=0\global\subsubsecno=0
\global\deno=0\global\prono=0\global\teno=0\eqnres@t\noindent
{\bf\the\secno. #1}
\writetoca{{\secsym} {#1}}\par\nobreak\medskip\nobreak}
\global\newcount\subsecno \global\subsecno=0
\def\subsec#1{\global\advance\subsecno by1\message{(\secsym\the\subsecno. #1)}
\ifnum\lastpenalty>9000\else\bigbreak\fi\global\subsubsecno=0
\global\deno=0\global\prono=0\global\teno=0
\noindent{\it\secsym\the\subsecno. #1}
\writetoca{\string\quad {\secsym\the\subsecno.} {#1}}
\par\nobreak\medskip\nobreak}
\global\newcount\subsubsecno \global\subsubsecno=0
\def\subsubsec#1{\global\advance\subsubsecno by1
\message{(\secsym\the\subsecno.\the\subsubsecno. #1)}
\ifnum\lastpenalty>9000\else\bigbreak\fi
\noindent\quad{\secsym\the\subsecno.\the\subsubsecno.}{#1}
\writetoca{\string\qquad{\secsym\the\subsecno.\the\subsubsecno.}{#1}}
\par\nobreak\medskip\nobreak}

\global\newcount\deno \global\deno=0
\def\de#1{\global\advance\deno by1
\message{(\bf Definition\quad\secsym\the\subsecno.\the\deno #1)}
\ifnum\lastpenalty>9000\else\bigbreak\fi
\noindent{\bf Definition\quad\secsym\the\subsecno.\the\deno}{#1}
\writetoca{\string\qquad{\secsym\the\subsecno.\the\deno}{#1}}}

\global\newcount\prono \global\prono=0
\def\pro#1{\global\advance\prono by1
\message{(\bf Proposition\quad\secsym\the\subsecno.\the\prono #1)}
\ifnum\lastpenalty>9000\else\bigbreak\fi
\noindent{\bf Proposition\quad\secsym\the\subsecno.\the\prono}{#1}
\writetoca{\string\qquad{\secsym\the\subsecno.\the\prono}{#1}}}

\global\newcount\teno \global\prono=0
\def\te#1{\global\advance\teno by1
\message{(\bf Theorem\quad\secsym\the\subsecno.\the\teno #1)}
\ifnum\lastpenalty>9000\else\bigbreak\fi
\noindent{\bf Theorem\quad\secsym\the\subsecno.\the\teno}{#1}
\writetoca{\string\qquad{\secsym\the\subsecno.\the\teno}{#1}}}
\def\subsubseclab#1{\DefWarn#1\xdef #1{\noexpand\hyperref{}{subsubsection}%
{\secsym\the\subsecno.\the\subsubsecno}%
{\secsym\the\subsecno.\the\subsubsecno}}%
\writedef{#1\leftbracket#1}\wrlabeL{#1=#1}}

\lockat

\def\np#1#2#3{Nucl. Phys. {\bf B#1} (#2) #3}
\def\pl#1#2#3{Phys. Lett. {\bf #1B} (#2) #3}

\def\physrev#1#2#3{Phys. Rev. {\bf D#1} (#2) #3}

\def\prep#1#2#3{Phys. Rep. {\bf #1} (#2) #3}

\def\IB{\relax\hbox{$\inbar\kern-.3em{\rm B}$}}
\def\IC{\relax\hbox{$\inbar\kern-.3em{\rm C}$}}
\def\ID{\relax\hbox{$\inbar\kern-.3em{\rm D}$}}
\def\IE{\relax\hbox{$\inbar\kern-.3em{\rm E}$}}
\def\IF{\relax\hbox{$\inbar\kern-.3em{\rm F}$}}
\def\IG{\relax\hbox{$\inbar\kern-.3em{\rm G}$}}
\def\IGa{\relax\hbox{${\rm I}\kern-.18em\Gamma$}}
\def\IH{\relax{\rm I\kern-.18em H}}
\def\IK{\relax{\rm I\kern-.18em K}}
\def\IL{\relax{\rm I\kern-.18em L}}
\def\IP{\relax{\rm I\kern-.18em P}}
\def\IR{\relax{\rm I\kern-.18em R}}
\def\IZ{\relax\ifmmode\mathchoice
{\hbox{\cmss Z\kern-.4em Z}}{\hbox{\cmss Z\kern-.4em Z}}
{\lower.9pt\hbox{\cmsss Z\kern-.4em Z}}
{\lower1.2pt\hbox{\cmsss Z\kern-.4em Z}}\else{\cmss Z\kern-.4em Z}\fi}

\def\II{\relax{\rm I\kern-.18em I}}

\def\CA {{\cal A}}

\def\CD {{\cal D}}
\def\CE {{\cal E}}
\def\CF {{\cal F}}

\def\CH {{\cal H}}

\def\CM {{\cal M}}

\def\CO {{\cal O}}
\def\CP {{\cal P}}

\def\CS {{\cal S}}


\def\p{\partial}
\def\pb{\bar{\partial}}

\def\zb {\bar{z}}


\def\Tr{{\rm Tr}}
\def\Id{{\rm Id}}

\def\p{\partial}
\def\pb{\bar{\partial}}

\def\lieg{{\underline{\bf g}}}

\def\inbar{\,\vrule height1.5ex width.4pt depth0pt}
\font\cmss=cmss10 \font\cmsss=cmss10 at 7pt


\font\manual=manfnt \def\dbend{\lower3.5pt\hbox{\manual\char127}}


\def\boxit#1{\vbox{\hrule\hbox{\vrule\kern8pt
\vbox{\hbox{\kern8pt}\hbox{\vbox{#1}}\hbox{\kern8pt}}
\kern8pt\vrule}\hrule}}
\def\mathboxit#1{\vbox{\hrule\hbox{\vrule\kern8pt\vbox{\kern8pt
\hbox{$\displaystyle #1$}\kern8pt}\kern8pt\vrule}\hrule}}

%
\lref\blzh{A. Belavin, V. Zakharov, ``Yang-Mills Equations as inverse
scattering
problem''Phys. Lett. B73, (1978) 53}
\lref\afs{Alekseev, Faddeev, Shatashvili,  }
\lref\bost{L. Alvarez-Gaume, J.B. Bost , G. Moore, P. Nelson, C.
Vafa,
``Bosonization on higher genus Riemann surfaces,''
Commun.Math.Phys.112:503,1987}
\lref\agmv{L. Alvarez-Gaum\'e,
C. Gomez, G. Moore,
and C. Vafa, ``Strings in the Operator Formalism,''
Nucl. Phys. {\bf 303}(1988)455}
\lref\atiyah{M. Atiyah, ``Green's Functions for
Self-Dual Four-Manifolds,'' Adv. Math. Suppl.
{\bf 7A} (1981)129}
\lref\dglsrev{M.~Douglas, 
``Superstring dualities, Dirichlet branes and the small scale structure of space'', hep-th/9610041}
\lref\sha{S.~Shatashvili, unpublished}

\lref\donagi{R.~Y.~ Donagi,
``Seiberg-Witten integrable systems'',
alg-geom/9705010 }

\lref\AHS{M.~ Atiyah, N.~ Hitchin and I.~ Singer, ``Self-Duality in
Four-Dimensional
Riemannian Geometry", Proc. Royal Soc. (London) {\bf A362} (1978)
425-461.}
\lref\fmlies{M. F. Atiyah and I. M. Singer,
``The index of elliptic operators IV,'' Ann. Math. {\bf 93}(1968)119}
\lref\BlThlgt{M.~ Blau and G.~ Thompson, ``Lectures on 2d Gauge
Theories: Topological Aspects and Path
Integral Techniques", Presented at the
Summer School in Hogh Energy Physics and
Cosmology, Trieste, Italy, 14 Jun - 30 Jul
1993, hep-th/9310144.}
\lref\bpz{A.A. Belavin, A.M. Polyakov, A.B. Zamolodchikov,
``Infinite conformal symmetry in two-dimensional quantum
field theory,'' Nucl.Phys.B241:333,1984}
\lref\braam{P.J. Braam, A. Maciocia, and A. Todorov,
``Instanton moduli as a novel map from tori to
K3-surfaces,'' Inven. Math. {\bf 108} (1992) 419}
\lref\CMR{ For a review, see
S. Cordes, G. Moore, and S. Ramgoolam,
`` Lectures on 2D Yang Mills theory, Equivariant
Cohomology, and Topological String Theory,''
Lectures presented at the 1994 Les Houches Summer School
 ``Fluctuating Geometries in Statistical Mechanics and Field
Theory.''
and at the Trieste 1994 Spring school on superstrings.
hep-th/9411210, or see http://xxx.lanl.gov/lh94}
\lref\dnld{S. Donaldson, ``Anti self-dual Yang-Mills
connections over complex  algebraic surfaces and stable
vector bundles,'' Proc. Lond. Math. Soc,
{\bf 50} (1985)1}
\lref\moorelitt{A. Losev, G. Moore, and 
S. Shatashvili, ``M$\&$m's,'' hep-th/9707250.}
 
\lref\natistring{N. Seiberg, ``New Theories in Six Dimensions and
Matrix Description of M-theory on $T^5$ and $T^5/\IZ_2$,''
hep-th/9705221.}%
 
\lref\helpol{S. Hellerman and J. Polchinski, ``Compactification in 
the Lightlike Limit,'' hep-th/9711037.}%
 
\lref\bfss{T. Banks, W. Fischler, S. Shenker, and L. Susskind, 
``M theory as a Matrix Model:  A Conjecture,'' hep-th/9610043,
\physrev{55}{1997}{112}.}%

\lref\DoKro{S.K.~ Donaldson and P.B.~ Kronheimer,
{\it The Geometry of Four-Manifolds},
Clarendon Press, Oxford, 1990.}
\lref\donii{
S. Donaldson, Duke Math. J. , {\bf 54} (1987) 231. }

\lref\fs{L. Faddeev and S. Shatashvili, Theor. Math. Fiz., 60 (1984)
206}
\lref\fsi{ L. Faddeev, Phys. Lett. B145 (1984) 81.}
\lref\fadba{L.D.~Faddeev, ``How  algebraic Bethe ansatz works for
integrable
model'',  hep-th/9605187}
\lref\nikitathes{N.~Nekrasov, PhD. Thesis, Princeton 1996\semi
``Five-dimensional gauge theories and
relativistic integrable systems'', hep-th/9609219}
\lref\fadbai{L.D.~Faddeev,
``Algebraic aspects of Bethe ansatz'',  Int.J.Mod.Phys.{\bf A}10 (1995)
1845-1878,
 hep-th/9404013 \semi
``The Bethe  ansatz'', SFB-288-70, Jun 1993. Andrejewski lectures}
\lref\faddeevlmp{L. D. Faddeev, ``Some Comments on Many Dimensional Solitons'',
Lett. Math. Phys., 1 (1976) 289-293.}

\lref\banksreview{T. Banks, ``Matrix Theory,'' hep-th/9710231.}%
 
\lref\lstrings{R.~Dijkgraaf, E.~Verlinde, H.~Verlinde, ``BPS spectrum of
the five-brane and black hole entropy'', hep-th/9603126\semi
R.~Dijkgraaf, E.~Verlinde, H.~Verlinde, ``BPS quantization 
of the five-brane'', hep-th/96040055\semi
J.~Schwarz, ``Self-dual string in six dimensions'', hep-th/9604171\semi
R.~Dijkgraaf, E.~Verlinde, H.~Verlinde, ``$5D$ black 
holes and matrix strings'' hep-th/9704018\semi
A.~Losev, G.~Moore, S.~Shatashvili, ``M $\&$ m's'', hep-th/9707250}
\lref\witthig{E.~Witten, ``On the conformal field theory of the Higgs branch'',
hep-th/9707093}
\lref\crt{
 N.~Seiberg, S.~Sethi, 
``Comments on Neveu-Schwarz Five-Branes'', hep-th/9708085
}
\lref\gerasimov{A. Gerasimov, ``Localization in
GWZW and Verlinde formula,'' hepth/9305090}

\lref\gottsh{L. Gottsche, Math. Ann. 286 (1990)193}
\lref\GrHa{P.~ Griffiths and J.~ Harris, {\it Principles of
Algebraic
geometry},
p. 445, J.Wiley and Sons, 1978. }

\lref\hitchin{N. Hitchin, ``Polygons and gravitons,''
Math. Proc. Camb. Phil. Soc, (1979){\bf 85} 465}
\lref\hi{N.~Hitchin, ``Stable bundles and integrable systems'', Duke Math
{\bf 54}  (1987),91-114}
\lref\hid{N.~Hitchin, ``The self-duality equations on a Riemann surface'',
Proc. London Math. Soc. {\bf 55} (1987) 59-126 }
\lref\hklr{N.~Hitchin, A.~Karlhede, U.~Lindstrom, and M.~Rocek,
``Hyperkahler metrics and supersymmetry,''
Commun. Math. Phys. {\bf 108}(1987)535}
\lref\hirz{F. Hirzebruch and T. Hofer, Math. Ann. 286 (1990)255}
\lref\btverlinde{M.~ Blau, G.~ Thomson,
``Derivation of the Verlinde Formula from Chern-Simons Theory and the
$G/G$
   model'',Nucl. Phys. {\bf B}408 (1993) 345-390 }
\lref\kronheimer{P. Kronheimer, ``The construction of ALE spaces as
hyper-kahler quotients,'' J. Diff. Geom. {\bf 28}1989)665}
\lref\kricm{P. Kronheimer, ``Embedded surfaces in
4-manifolds,'' Proc. Int. Cong. of
Math. (Kyoto 1990) ed. I. Satake, Tokyo, 1991}

\lref\krmw{P.~Kronheimer and T.~Mrowka,
``Gauge theories for embedded surfaces I,''
Topology {\bf 32} (1993) 773,
``Gauge theories for embedded surfaces II,''
preprint.}
\lref\kirwan{F.~Kirwan, ``Cohomology of quotients in symplectic
and algebraic geometry'', Math. Notes, Princeton University Press, 1985}
\lref\avatar{A. Losev, G. Moore, N. Nekrasov, S. Shatashvili,
``Four-Dimensional Avatars of 2D RCFT,''
hep-th/9509151, Nucl.Phys.Proc.Suppl.46:130-145,1996 }
\lref\cocycle{A. Losev, G. Moore, N. Nekrasov, S. Shatashvili,
``Central Extensions of Gauge Groups Revisited,''
hep-th/9511185.}
\lref\maciocia{A. Maciocia, ``Metrics on the moduli
spaces of instantons over Euclidean 4-Space,''
Commun. Math. Phys. {\bf 135}(1991) , 467}
\lref\mickold{J. Mickelsson, CMP, 97 (1985) 361.}
\lref\mick{J. Mickelsson, ``Kac-Moody groups,
topology of the Dirac determinant bundle and
fermionization,'' Commun. Math. Phys., {\bf 110} (1987) 173.}
\lref\milnor{J. Milnor, ``A unique decomposition
theorem for 3-manifolds,'' Amer. Jour. Math, (1961) 1}
\lref\taming{G. Moore and N. Seiberg,
``Taming the conformal zoo,'' Phys. Lett.
{\bf 220 B} (1989) 422}
\lref\nair{V.P.Nair, ``K\"ahler-Chern-Simons Theory'', hep-th/9110042}
\lref\ns{V.P. Nair and Jeremy Schiff,
``Kahler Chern Simons theory and symmetries of
antiselfdual equations'' Nucl.Phys.B371:329-352,1992;
``A Kahler Chern-Simons theory and quantization of the
moduli of antiselfdual instantons,''
Phys.Lett.B246:423-429,1990,
``Topological gauge theory and twistors,''
Phys.Lett.B233:343,1989}
\lref\ogvf{H. Ooguri and C. Vafa, ``Self-Duality
and $N=2$ String Magic,'' Mod.Phys.Lett. {\bf A5} (1990) 1389-1398;
``Geometry
of$N=2$ Strings,'' Nucl.Phys. {\bf B361}  (1991) 469-518.}
\lref\park{J.-S. Park, ``Holomorphic Yang-Mills theory on compact
Kahler
manifolds,'' hep-th/9305095; Nucl. Phys. {\bf B423} (1994) 559;
J.-S.~ Park, ``$N=2$ Topological Yang-Mills Theory on Compact
K\"ahler
Surfaces", Commun. Math, Phys. {\bf 163} (1994) 113;
S. Hyun and J.-S.~ Park, ``$N=2$ Topological Yang-Mills Theories and Donaldson
Polynomials", hep-th/9404009}
\lref\parki{S. Hyun and J.-S. Park,
``Holomorphic Yang-Mills Theory and Variation
of the Donaldson Invariants,'' hep-th/9503036}
\lref\dpark{J.-S.~Park, ``Monads and D-instantons'', hep-th/9612096}
\lref\pohl{Pohlmeyer, Commun.
Math. Phys. {\bf 72}(1980)37}
\lref\pwf{A.M. Polyakov and P.B. Wiegmann,
Phys. Lett. {\bf B131}(1983)121}
\lref\clash{
A.~Losev, G.~Moore, N.~Nekrasov, S.~Shatashvili,
`` Chiral  Lagrangians, Anomalies, Supersymmetry, and Holomorphy'', Nucl.Phys.
{\bf B} 484(1997) 196-222, hep-th/9606082 }
\lref\givental{A.B.~Givental,
``Equivariant Gromov - Witten Invariants'',
alg-geom/9603021}
\lref\thooft{G. 't Hooft , ``A property of electric and
magnetic flux in nonabelian gauge theories,''
Nucl.Phys.B153:141,1979}
\lref\vafa{C. Vafa, ``Conformal theories and punctured
surfaces,'' Phys.Lett.199B:195,1987 }
\lref\adhm{M. Atiyah, V. Drinfeld, N. Hitchin, and Y. Manin,
``Construction of Instantons,'' \pl{65}{1978}{185}.}

\lref\vrlsq{E. Verlinde and H. Verlinde,
``Conformal Field Theory and Geometric Quantization,''
in {\it Strings'89},Proceedings
of the Trieste Spring School on Superstrings,
3-14 April 1989, M. Green, et. al. Eds. World
Scientific, 1990}

\lref\mwxllvrld{E. Verlinde, ``Global Aspects of
Electric-Magnetic Duality,'' hep-th/9506011}

\lref\wrdhd{R. Ward, Nucl. Phys. {\bf B236}(1984)381}
\lref\ward{Ward and Wells, {\it Twistor Geometry and
Field Theory}, CUP }


\lref\WitDonagi{R.~ Donagi, E.~ Witten,
``Supersymmetric Yang-Mills Theory and
Integrable Systems'', hep-th/9510101, Nucl.Phys.{\bf B}460 (1996) 299-334}
\lref\Witfeb{E.~ Witten, ``Supersymmetric Yang-Mills Theory On A
Four-Manifold,'' J. Math. Phys. {\bf 35} (1994) 5101.}
\lref\Witr{E.~ Witten, ``Introduction to Cohomological Field
Theories",
Lectures at Workshop on Topological Methods in Physics, Trieste, Italy,
Jun 11-25, 1990, Int. J. Mod. Phys. {\bf A6} (1991) 2775.}
\lref\Witgrav{E.~ Witten, ``Topological Gravity'', Phys.Lett.206B:601, 1988}
\lref\witaffl{I. ~ Affleck, J.A.~ Harvey and E.~ Witten,
	``Instantons and (Super)Symmetry Breaking
	in $2+1$ Dimensions'', Nucl. Phys. {\bf B}206 (1982) 413}
\lref\wittabl{E.~ Witten,  ``On $S$-Duality in Abelian Gauge Theory,''
hep-th/9505186; Selecta Mathematica {\bf 1} (1995) 383}
\lref\wittgr{E.~ Witten, ``The Verlinde Algebra And The Cohomology Of
The Grassmannian'',  hep-th/9312104}
\lref\wittenwzw{E. Witten, ``Nonabelian bosonization in
two dimensions,'' Commun. Math. Phys. {\bf 92} (1984)455 }
\lref\witgrsm{E. Witten, ``Quantum field theory,
grassmannians and algebraic curves,'' Commun.Math.Phys.113:529,1988}
\lref\wittjones{E. Witten, ``Quantum field theory and the Jones
polynomial,'' Commun.  Math. Phys., 121 (1989) 351. }
\lref\witttft{E.~ Witten, ``Topological Quantum Field Theory",
Commun. Math. Phys. {\bf 117} (1988) 353.}
\lref\wittmon{E.~ Witten, ``Monopoles and Four-Manifolds'', hep-th/9411102}
\lref\Witdgt{ E.~ Witten, ``On Quantum gauge theories in two
dimensions,''
Commun. Math. Phys. {\bf  141}  (1991) 153\semi
 ``Two dimensional gauge
theories revisited'', J. Geom. Phys. 9 (1992) 303-368}
\lref\Witgenus{E.~ Witten, ``Elliptic Genera and Quantum Field Theory'',
Comm. Math. Phys. 109(1987) 525. }
\lref\OldZT{E. Witten, ``New Issues in Manifolds of SU(3) Holonomy,''
{\it Nucl. Phys.} {\bf B268} (1986) 79 \semi
J. Distler and B. Greene, ``Aspects of (2,0) String Compactifications,''
{\it Nucl. Phys.} {\bf B304} (1988) 1 \semi
B. Greene, ``Superconformal Compactifications in Weighted Projective
Space,'' {\it Comm. Math. Phys.} {\bf 130} (1990) 335.}

\lref\bagger{E.~ Witten and J. Bagger, Phys. Lett.
{\bf 115B}(1982) 202}

\lref\witcurrent{E.~ Witten,``Global Aspects of Current Algebra'',
Nucl.Phys.B223 (1983) 422\semi
``Current Algebra, Baryons and Quark Confinement'', Nucl.Phys. B223 (1993)
433}
\lref\Wittreiman{S.B. Treiman,
E. Witten, R. Jackiw, B. Zumino, ``Current Algebra and
Anomalies'', Singapore, Singapore: World Scientific ( 1985) }
\lref\Witgravanom{L. Alvarez-Gaume, E.~ Witten, ``Gravitational Anomalies'',
Nucl.Phys.B234:269,1984. }

\lref\CHSW{P.~Candelas, G.~Horowitz, A.~Strominger and E.~Witten,
``Vacuum Configurations for Superstrings,'' {\it Nucl. Phys.} {\bf
B258} (1985) 46.}

\lref\AandB{E.~Witten, in ``Proceedings of the Conference on Mirror Symmetry",
MSRI (1991).}

\lref\wittenwzw{E. Witten, ``Nonabelian bosonization in
two dimensions,'' Commun. Math. Phys. {\bf 92} (1984)455 }
\lref\grssmm{E. Witten, ``Quantum field theory,
grassmannians and algebraic curves,'' Commun.Math.Phys.113:529,1988}
\lref\wittjones{E. Witten, ``Quantum field theory and the Jones
polynomial,'' Commun.  Math. Phys., 121 (1989) 351. }
\lref\wittentft{E.~ Witten, ``Topological Quantum Field Theory",
Commun. Math. Phys. {\bf 117} (1988) 353.}
\lref\Witdgt{ E.~ Witten, ``On Quantum gauge theories in two
dimensions,''
Commun. Math. Phys. {\bf  141}  (1991) 153.}
\lref\Witfeb{E.~ Witten, ``Supersymmetric Yang-Mills Theory On A
Four-Manifold,'' J. Math. Phys. {\bf 35} (1994) 5101.}
\lref\Witr{E.~ Witten, ``Introduction to Cohomological Field
Theories",
Lectures at Workshop on Topological Methods in Physics, Trieste,
Italy,
Jun 11-25, 1990, Int. J. Mod. Phys. {\bf A6} (1991) 2775.}
\lref\wittabl{E. Witten,  ``On S-Duality in Abelian Gauge Theory,''
hep-th/9505186}

\lref\seiken{K. Intriligator, N. Seiberg,
``Mirror Symmetry in Three Dimensional Gauge Theories'',
hep-th/9607207, Phys.Lett. B387 (1996) 513}
\lref\douglas{M.R. Douglas, ``Enhanced Gauge
Symmetry in M(atrix) Theory,'' hep-th/9612126}
\lref\hs{J.A. Harvey and A. Strominger,
``The heterotic string is a soliton,''
hep-th/9504047}
\lref\hm{ J.A.~Harvey, G.~Moore,
``On the algebras of BPS states'', hep-th/9609017}
\lref\abs{O.~Aharony, M.~Berkooz, N.~Seiberg, ``Light-cone description
of $(2,0)$ supersonformal theories in six dimensions'', hep-th/9712117}
\lref\sen{A. Sen, `` String- String Duality Conjecture In Six Dimensions And
Charged Solitonic Strings'',  hep-th/9504027}
\lref\KN{P.~Kronheimer and H.~Nakajima,  ``Yang-Mills instantons
on ALE gravitational instantons,''  Math. Ann.
{\bf 288}(1990)263}
\lref\nakajima{H.~Nakajima, ``Homology of moduli
spaces of instantons on ALE Spaces. I'' J. Diff. Geom.
{\bf 40}(1990) 105; ``Instantons on ALE spaces,
quiver varieties, and Kac-Moody algebras,'' Duke. Math. J. {\bf 76} (1994)
365-416\semi
``Gauge theory on resolutions of simple singularities
and affine Lie algebras,'' Inter. Math. Res. Notices (1994), 61-74}
\lref\nakheis{H.~Nakajima, ``Heisenberg algebra and Hilbert schemes of
points on
projective surfaces'',  alg-geom/9507012\semi
``Lectures on Hilbert
 schemes of points on surfaces'', H.~Nakajima's
homepage}
\lref\vw{C.~Vafa, E.~Witten, ``A strong coupling test of $S$-duality'',
Nucl. Phys. {\bf B} 431 (1994) 3-77}
\lref\grojn{I. Grojnowski, ``Instantons and
affine algebras I: the Hilbert
 scheme and
vertex operators,'' alg-geom/9506020}
\lref\dvafa{C.~Vafa, ``Instantons on D-branes'', hep-th/9512078,
Nucl.Phys. B463 (1996) 435-442}
\lref\atbotti{M.~Atiyah, R.~Bott, ``The Yang-Mills Equations Over
Riemann Surfaces'', Phil. Trans. R.Soc. London A {\bf 308}, 523-615 (1982)}

\lref\naka{H.~Nakajima, ``Resolutions of Moduli Spaces of Ideal Instantons on $\IR^{4}$'',
in ``Topology, Geometry and Field Theory'', eds. Fukaya, Furuta, Kohno and Kotschick, World Scientific}

\lref\witbound{E.~Witten, ``Bound states of strings and $p$-branes'', 
\np{460}{1990}{335}, hep-th/9510135} 
\lref\connes{A.~Connes, ``Non-commutative Geometry'', Academic Press, 1994}
\lref\sbs{R.~Leigh and M.~Rozali, hep-th/9712168\semi P.-M.~Ho, Y.-Y.~Wu
and Y.-S.~Wu, hep-th/9712201\semi P.-M.~Ho, Y.-S.~Wu, hep-th/9801147\semi
M.~Li, hep-th/9802052 }

\Title{ \vbox{\baselineskip12pt\hbox{hep-th/9802068}
\hbox{ITEP-TH-9/98}
\hbox{HUTP- 98/A004}
\hbox{Davis-02/98}}}
{\vbox{
 \centerline{Instantons on noncommutative $\IR^{4}$,}
 \centerline{and $(2,0)$ superconformal six dimensional theory}}}
\medskip
\centerline{Nikita Nekrasov $^1$, Albert Schwarz $^2$}

\vskip 0.5cm
\centerline{$^{1}$ Institute of Theoretical and Experimental
Physics,
117259, Moscow, Russia}
\centerline{$^1$ Lyman Laboratory of Physics,
Harvard University, Cambridge, MA 02138}
\centerline{$^{2}$ Department of Mathematics, UCDavis, Davis, CA 95616}
\vskip 0.1cm

\centerline{nikita@string.harvard.edu }
\centerline{schwarz@math.ucdavis.edu}

\medskip
\noindent
We show that the resolution of moduli space of ideal instantons  parameterizes the
instantons
on non-commutative $\IR^{4}$. This moduli space appears as a Higgs branch of
the theory of $k$ $D0$-branes bound to $N$ $D4$-branes by the expectation value of the 
$B$ field.  It also appears as a regularized version of the target space of supersymmetric
quantum mechanics arising in the light cone description of 
$(2,0)$ superconformal theories in six dimensions.

\Date{02/98}

\newsec{Introduction }

The appearence of non-commutative geometry \connes\  in the  
physics of $D$-branes has been
anticipated with the very 
understanding of the fact that the gauge theory on the
worldvolume of $N$ coincident $D$-branes is a non-abelian gauge theory 
\witbound\dglsrev. In this theory
the scalar fields $X_{i}$  in the adjoint representation are the
non-abelian generalizations of the transverse coordinates of the branes. 
\lref\brs{
M. Rozali, ``Matrix Theory and U Duality in Seven Dimensions,''
hep-th/9702136, \pl{400}{1997}{260}\semi
M. Berkooz, M. Rozali and N. Seiberg,  ``Matrix Description
of M theory on $T^4$ and $T^5$,'' hep-th/9704089, \pl{408}{1997}{105}.}%

It is known that the compactifications of Matrix theory \bfss\banksreview\
 on tori 
${\bf T}^{d}$ exhibit richer structures as
the dimensionality $d$ of the tori increases \brs\natistring.  The 
compactification on a torus implies that the certain 
constraints are imposed on the matrices $X_{i}$:
\eqn\comp{
X_{i} + 2\pi R_{i} \delta_{ij} = U_{j} X_{i} U_{j}^{-1},}
It seems natural 
to study all possible solutions $U_{i}$ to the consistency equations 
for the compactification of the matrix fields\foot{The conjecture of \sha\ is that the non-abelian
tensor fields in six dimensions would also 
appear as such solutions} \sha. 

Recently, 
the  non-commutative torus emerged as one of the solutions to \comp\
\lref\cds{A. Connes, M. R. Douglas and A. Schwarz, ``Noncommutative
Geometry and Matrix Theory: Compactification on Tori,'' hep-th/9711162.}%
\cds. 
It has been argued that the parameter of non-commutativity is related to the flux
of the $B$-field through the torus. 
\lref\dh{M.~Douglas, C.~Hull, ``$D$-branes and non-commutative geometry'', hep-th/9711165}
It has been further shown in \dh\ that the compactification on a non--commutative torus can be thought of  as 
a $T$-dual to a 
limit of the conventional compactification on a commutative torus.
See \sbs\ for further developments in the studies of compactifications on 
low-dimensional
tori.

On the other hand, the modified 
self-duality equations on the matrices in the Matrix description of fivebrane
theory has been used in \abs\ in the study of quantum mechanics on 
the instanton moduli space.
The modification is most easily described in the framework of ADHM equations. It makes the
moduli space smooth and allows to define a six dimensional theory decoupled
from the eleven-dimensional supergravity and all others $M$-theoretic degrees of freedom.
The heuristric reason for the possibility of such decoupling
is the fact that the Higgs branch of the theory is smooth and there is
no place for the Coulomb branch to touch it.  

In this paper we propose an explanation of the latter construction in terms
of non-commutative geometry. We show, that the solutions
to modified ADHM equations parameterize (anti-)self-dual gauge fields on non-commutative
$\IR^{4}$.

\newsec{Instantons on a commutative space}

Let $X$ be a four dimensional compact Riemannian manifold and $P$ a principal $G$-bundle on it,
with $G = U(N)$. The connection $A$ is called anti-self-dual (ASD), or instanton, if its curvature obeys the equation:
$$
F^{+} := {\half} ( F + * F) = 0
$$
where $*: \Omega^{k} \to \Omega^{4-k}$ is the Hodge star. 
The importance of ASD gauge fields in physics stems from the fact that they minimize the Yang-Mills
action in a given topological sector, i.e. for fixed
$k = -{1\over{8\pi^{2}}} \int {\Tr} F \wedge F$. In supersymmetric gauge theories the instantons 
are the configurations of 
gauge fields which preserve some supersymmetry, since the self-dual part of $F$ appears in the right hand side of susy
transformations. For the same reason they play a major role in Matrix theory.

The space of ASD gauge gauge fields modulo gauge transformations is called the moduli
space of instantons $\CM_{k}$ and in a generic situation it is a smooth
 manifold
of dimension 
$$
4Nk - {{N^{2}-1}\over{2}}  ({\chi} + {\sigma})
$$ 
where $\chi$ and $\sigma$ are the Euler characteristics and signature of $X$ respectively.

The moduli space $\CM$ is non-compact. The lack of compactness is due to the
so-called point-like instantons. What can happen is that for a sequence of ASD connections
$A_{i}$ the region $D_{i}$ where some topological charge 
$-{1\over{8\pi^{2}}} \int_{D_{i}} {\Tr} F \wedge F$ is concentrated can
 shrink to a zero size. There exists
\lref\uhl{K.~Uhlenbeck, ``Removable Singularities in Yang-Mills Fields'', 
Com.. Math. Phys. {\bf 83} (1982) 11}
a compactification $\bar\CM_{k}$ due to Uhlenbeck \uhl\ which simply adds the 
centers of the point-like instantons:
$$
\bar\CM_{k} = \CM_{k} \cup \CM_{k-1} \times X \cup \ldots \cup
\CM_{k-l} \times {\rm Sym}^{l}X \ldots
$$
which is suitable for certain purposes but not for all. In particular, the space $\bar\CM_{k}$ has
orbifold singularities. 

One can also study non-compact spaces, $X = \IR^{4}$ being the first example. In posing a moduli
problem
one has to specify the conditions on the behavior of the gauge fields at infinity. The natural condition
is : $A_{\mu} \sim g^{-1}\p_{\mu} g  + O({1\over{r^{2}}})$ as $r \to \infty$. 
One may also restrict the allowed gauge transformations to those which tend to $1$ at infinity.

\subsec{Review of ADHM construction}

Atiyah-Drinfeld-Hitchin-Manin 
describe \adhm\  a way of getting the solutions obeying the asymptotics stated
above to the instanton equations on $\IR^{4}$ 
in terms of solutions to some quadratic matrix equations. More specifically, in order
to describe charge $N$ instantons with gauge group $U(k)$ one starts with the following data:

\item{1.} A pair of complex hermitian 
vector spaces $V = \IC^{N}$ and $W= \IC^{k}$;

\item{2.} The operators $B_{0}, B_{1} \in Hom (V, V)$, $I \in Hom (W, V)$, $J \in Hom (V, W)$;

which must obey the equations $\mu_{r} = 0, \mu_{c} = 0$, where:

\eqn\adhme{\eqalign{\mu_{r} = & [B_{0}, B_{0}^{\dagger}] + [ B_{1}, B_{1}^{\dagger}] + II^{\dagger} - J^{\dagger}J\cr
\mu_{c} = & [ B_{0}, B_{1} ] + IJ \cr}}

There is also a non-degeneracy condition which must be imposed by hand, namely,
the set $(B, I, J)$ should have trivial stabilizer in the $U(V)$ group.

For $z = (z_{0}, z_{1}) \in \IC^{2} \approx \IR^{4}$ define an operator 
$\CD^{\dagger}_{z} : V \oplus V \oplus W \to V \oplus V$ by the formula:
\eqn\de{\eqalign{{\CD}^{\dagger}_{z} = & \pmatrix{\tau_{z} \cr \sigma_{z}^{\dagger}\cr}, \cr
\tau_{z} = \pmatrix{B_{0} - z_{0} & -B_{1} + z_{1} & I\cr} \quad & \quad 
\sigma_{z} = \pmatrix{B_{1} - z_{1} \cr B_{0} - z_{0} \cr J \cr}\cr}}

Given the matrices obeying all the conditions above the actual instanton solution is 
determined by the following rather explicit formulae:

\eqn\instfl{A_{\mu} = \psi^{\dagger} \p_{\mu} \psi}
where $\psi : W \to V \oplus V \oplus W$ solves the equations:
$\CD^{\dagger} \psi = 0$ and is normalized : $\psi^{\dagger}\psi = 1$.

The moduli space of instantons with fixed framing at inifnity is identified with 
\eqn\om{\CM = \left( \mu_{r}^{-1}(0) \cap \mu_{c}^{-1}(0) \right) / U(V)}
where $U(V)$ is the group of unitary transformations of $V$ acting on the matrices $B,I,J$ in a natural 
way. 

\subsec{Regularization of ADHM data}

As we noted above the compactification $\bar\CM_{k}$ is a singular manifold. One may resolve it
to a smooth
 variety by 
 deforming the equations $\mu_{r}=\mu_{c}=0$
to 
$\mu_{r} = \zeta_{r} \Id, \quad \mu_{c} = 0$. 
One may add a constant to $\mu_{c}$
as well but this modification is equivalent to the one already considered by a linear
transformation of the data $B_{0,1}, B_{0,1}^{\dagger}, I, I^{\dagger}, J, J^{\dagger}$. 
The modification has been studied mathematically by various people
and we recommend the beautiful lectures by H.~Nakajima \nakheis\ for a review.
The deformed data form a moduli space:
\eqn\nmdl{\CM_{\zeta} = \left( \mu_{r}^{-1}(\zeta {\Id} ) \cap \mu_{c}^{-1}(0) \right) / U(V)} 
and they parameterize the torsion free sheaves on $\IC\IP^{2}$ with fixed
framing at the line at infinity. 
As we shall show in the next section, the deformed moduli space parameterizes the instantons on the
non-commutative $\IR^{4}$. 

\newsec{Instantons on non-commutative spaces}

The paradigm of non-commutative geometry is to describe the geometry of
ordinary spaces in terms of the algebra of (smooth, continious,..) functions and then
generalizing to the non-commutative case. In this sense, the non-commutative
$\IR^{4}$ is the algebra $\CA_{\omega}$ generated by 
$x_{\alpha}, \alpha = 1,2,3,4$ which obey the relations:
\eqn\dfnrl{[ x_{\alpha}, x_{\beta} ] = \omega_{\alpha\beta}}
where $\omega_{\alpha\beta}$ is a constant antisymmetric matrix. There are three
distinct cases one may consider:

\item{1.} $\omega$ has rank $0$. In this case $\CA_{\omega}$ is isomorphic to the algebra of functions on
the ordinary $\IR^{4}$. 

\item{2.} $\omega$ has rank $2$. In this case $\CA_{\omega}$ is the algebra of functions on the 
ordinary $\IR^{2}$ times the non-commutative $\IR^{2} = \{ (p,q) \vert [p,q] = -i\}$.

\item{3.} $\omega$ has rank $4$. 
In this case $\CA_{\omega}$ is {\it the } non-commutative $\IR^{4}$. Let $\pi^{\alpha\beta}$ be the inverse matrix to $\omega_{\alpha\beta}$ and 
$x^{\alpha} = \pi^{\alpha\beta} x_{\beta}$. Let $\vec x = ( x^{\alpha})$
and $\vec x^{\vee} = ( x_{\alpha})$.

The algebra depends essentially on one 
number $\zeta$ (which can be scaled away, but we shall keep
it in order to able to take a limit to the commutative case). We shall denote it as $\CA_{\zeta}$. Introduce the generators:
$z_{0} = x^{1} +  i x^{2}, z_{1} = x^{3} + i x^{4}$, then\foot{So,
$\pi^{\mu\nu}$ is proportional to $\zeta$, while $\omega_{\alpha\beta}$
is proportional to ${1\over{\zeta}}$ in accordance
with the standard quasiclassical limits}
\eqn\ncrf{[z_{0} , \zb_{0}] = [z_{1}, \zb_{1}] = -{\zeta \over {2}}}
The commutation relations \ncrf\ have an obvious group of 
authomorphismes of the form
 $x_{\alpha} \mapsto x_{\alpha} + \beta_{\alpha} \cdot 1$,
$\beta_{\alpha} \in \IR$. We denote the Lie algebra of this group by
$\lieg$. 
For the algebra \ncrf\ to represent the algebra of real-valued 
functions (and be represented by Hermitian operators) we need 
$\zeta \in \IR$. ( In the language of mathematics we can say that 
this condition means that the algebra at hand has an involution.) We
choose $\zeta >0$.  Of course, the algebra of polynomials in $z, \zb$
should be completed in some way. We propose to start with the 
algebra ${\rm End} {\CH}$ of
operators acting in the Fock space $\CH = \sum_{(n_{0}, n_{1}) \in 
\IZ_{+}^{2}} \IC \vert 
n_{0}, n_{1} \rangle$ where $z, \zb$ are represented as creation-annihilation
operators:
\eqn\rprs{\matrix{
 z_{0} \vert n_{0}, n_{1} \rangle = 
{\zeta \over{2}} \sqrt{n_{0}+1} \vert n_{0} +1 , n_{1} \rangle &  
\zb_{0} \vert n_{0}, n_{1} \rangle = {\zeta \over{2}} \sqrt{n_{0}} 
\vert n_{0} - 1 , n_{1} \rangle \cr
z_{1} \vert n_{0}, n_{1} \rangle = 
{\zeta \over{2}} \sqrt{n_{1}+1} \vert n_{0} , n_{1} +1 \rangle &  
\zb_{1} \vert n_{0}, n_{1} \rangle = {\zeta \over{2}} \sqrt{n_{1}} 
\vert n_{0}, n_{1} - 1 \rangle \cr}}
The algebra  ${\rm End}{\CH}$ has a subalgebra ${\rm End}_{0}\CH$
 of operators $A$ which have
finite norm; we will take the Hilbert-Schmidt norm:
${\Tr}_{\CH}(AA^{\dagger})^{\half}$.
We consider  an algebra $\CA_{\zeta}$ 
 defined as a subalgebra of ${\rm End}_{0}{\CH}$
which consists of smooth operators, i.e. those $A$ for which the
function $f_{A}: \lieg \to {\rm End}_{0}\CH$, $f_{A}(\vec t) = Ad_{\vec t \cdot
\vec x} A$ is smooth.    
Notice that $1$ does not belong to this algebra. This is 
a consequence of non-compactness of $\IR^{4}$. 
\lref\rieff{M.~Rieffel, ``Deformation quantization for actions of
$\IR^{d}$'', Memoirs of AMS, vol. 106, N 506 (1993)}
One can represent the elements of the Fock space $\CH$ as $L^{2}$
functions in 
two variables $q_{0} = x^{1}$ and $q_{1} = x^{3}$ (oscillator
wave functions in  real polarization).
It is actually possible to prove using the results of \rieff\ that 
smooth operators are those whose matrix elements ${\bf A}(q_{0}^{\prime},
q_{1}^{\prime}; q_{0}, q_{1})$ belong to the Schwartz space $\CS (\IR^{2} 
\times
\IR^{2} )$.
We consider representations of the involutive algebra $\CA_{\zeta}$
by means of operators in Hilbert space ( Hilbert modules with involution 
in mathematical
terminology).
A vector bundle over a non-commutative space $\CA$ is a projective 
module $\CE$, i.e. such 
a  module  for which another module $\CE^{\perp}$ exists with the property
$\CE \oplus \CE^{\perp} = \CA \oplus \ldots \oplus \CA$ 
(The last module is called free. In commutative
case free modules correspond to trivial bundles). 
In our description of
instantons over $\IR^{4}$ we shall be dealing with free modules only.

The notion of connection in the bundle $\CE$ has several definitions. 
The most convenient for us
at the moment will be the following one. Let $\lieg \subset {\rm Aut}({\CA})$
be a Lie sub-algebra of the algebra of authomorphismes of $\CA$. 
Then the connection $\nabla$ is an operator
$\nabla: \lieg \times \CE \to \CE$ which obeys the Leibnitz rule:
\eqn\rle{\nabla_{\xi} ( f \cdot s) = f \cdot \nabla_{\xi}s + {\xi}
 ( f) \cdot s, \qquad f \in \CA, \quad \xi \in \lieg,  \quad s \in \CE}
The curvature of $\nabla$ is an operator:
\eqn\crve{F({\nabla}) : \Lambda^{2}\lieg \times \CE \to \CE, \quad F({\nabla}) (\xi, \eta) = [\nabla_{\xi}, \nabla_{\eta}] - 
\nabla_{[\xi, \eta]}}

\subsec{Instantons on non-commutative $\IR^{4}$}

Just like in the commutative case one may define the Yang-Mills action, the instanton equations etc. See \connes\ for the definitions and \cds\ for recent discussion. 
We are looking at the solutions of the ASD conditions
$$
F^{+} = 0
$$
on the connection in the module $\CE$ over $\CA_{\zeta}$, where the $+$ sign means the self-dual part
with respect to the natural Hodge star operator acting on the $\Lambda^{2}\lieg$ where $\lieg \approx \IR^{4}$
is the abelian Lie algebra of authomorphismes of $\CA_{\zeta}$.

Now we are going to show that the resolution \nmdl\ gives rise to the ADHM description of
instantons on the non-commutative $\IR^{4}$.

Indeed, the core of the ADHM construction are the equations
\eqn\adhm{
\tau_{z} \sigma_{z} = 0,\quad  \tau_{z}\tau_{z}^{\dagger} = \sigma_{z}^{\dagger}\sigma_{z}}
where the operators $\sigma_{z}, \tau_{z}$ are constructed as above. 
Now suppose that the matrices $B, B^{\dagger},I^{\dagger}, I,J, J^{\dagger}$ 
obey the modified equations \nmdl. Then 
\adhm\ are no longer valid but they will be valid if the 
coordinate functions $z_{i}, {\zb}_{i}$ will not commute!
In fact, by imposing the commutation law 
$[z_{0}, {\zb}_{0} ] = [ z_{1}, {\zb}_{1} ] = - {{\zeta}\over{2}}$
we fulfill \adhm\  as the term with $\zeta$ from commutators
of $B$'s is now compensated
by the commutators of $z$'s! We now follow the steps of the ordinary ADHM
construction. We define an operator:
\eqn\ddg{\CD^{\dagger}_{z}: \left( V \oplus V \oplus W \right) \otimes \CA_{\zeta}
\to \left( V \oplus V \right) \otimes \CA_{\zeta}}
by the same formula \de. We look for the solution to the equation
\eqn\drdc{\CD^{\dagger}_{z}\psi = 0, \quad \psi : W\otimes \CA_{\zeta} \to 
 \left( V \oplus V \oplus W \right) \otimes \CA_{\zeta}} which is again normalized: $\psi^{\dagger}\psi = \Id_{W\otimes \CA_{\zeta}}$. These are defined up to unitary gauge transformations $g$ acting on $\psi$ on the right.  
Again, $A_{\mu} 
= \psi^{\dagger} \p_{\mu} \psi$, where the derivative is
understood as the action of $\lieg = \IR^{4}$ on $\CA_{\zeta}$ by translations.
We may now derive the formula for the curvature of the gauge field $A$. 
The derivation is very simlar
to the commutative case and it yields:
\eqn\crvt{F = \psi^{\dagger} \left( d \CD_{z} {1\over{\CD_{z}^{\dagger}\CD_{z}}} d \CD^{\dagger}_{z} \right) \psi}

The operator $\CD^{\dagger}_{z}\CD_{z} : 
( V \oplus V ) \otimes \CA_{\zeta} \to (V \oplus V) \otimes \CA_{\zeta}$ 
is again block-diagonal:
\eqn\lpl{\CD_{z}^{\dagger}\CD_{z} = 
\pmatrix{\Delta_{z} & 0\cr 0 & \Delta_{z} \cr}, \quad \Delta_{z} = 
\tau_{z}\tau_{z}^{\dagger}=
\sigma_{z}^{\dagger}\sigma_{z}}
Hence, just as in the commutative case:
\eqn\crvti{\eqalign{F = & \psi^{\dagger} \left( d\tau_{z}^{\dagger} {1\over{\Delta_{z}}} d\tau_{z} + 
d\sigma_{z} {1\over{\Delta_{z}}} d\sigma_{z}^{\dagger}\right) \psi = \cr
& = \psi^{\dagger} 
\pmatrix{
d{\zb}_{0} {1\over{\Delta}_{z}} dz_{0}  + dz_{1} 
{1\over{\Delta}_{z}} d{\zb}_{1}
& - d{\zb}_{0} {1\over{\Delta}_{z}} dz_{1} + dz_{1} {1\over{\Delta}_{z}} d{\zb}_{0}& 0 \cr 
   - d{\zb}_{1} {1\over{\Delta}_{z}} dz_{0} + dz_{0} {1\over{\Delta}_{z}} d{\zb}_{1}
 &   d{\zb}_{1} {1\over{\Delta}_{z}} dz_{1}  + dz_{0} {1\over{\Delta}_{z}} d{\zb}_{0}& 0\cr 0 & 0 & 0 \cr}\psi\cr}}
which is anti-self-dual. We should warn the reader that
here $dz_{i}, d\zb_{i}$ are just the generating anti-commuting parameters for representing the matrix $F_{\nabla}(\xi,\eta)$
in  short print. They commute with $z_{i}, \zb_{i}$. 

\subsec{Commutative interpretation of the non-commutative equations}

The construction above is not yet completely rigorous. We must prove that
$\psi$ exists and that  $\Delta_{z}$ is invertible. A useful technique is to
represent the equations over $\CA_{\zeta}$ in terms of the
(perhaps differential) equations on ordinary functions. 
In this way the multiplication of operators $a \cdot b$ is mapped to 
\eqn\myl{a \star b (x) = e^{{\half} \pi^{\mu\nu} {{\p^{2}}\over{\p \xi^{\mu}\p 
\eta^{\nu}}} } a(x + \xi) b(x + \eta) \vert_{\xi = \eta =0}  }

Now we can study the questions we posed in the beginning of the section.

 It follows
from \lpl\ that 
the
condition that $\upsilon$ is a zero mode of $\Delta_{z}$ is 
equivalent to the conditions:
$\tau_z^{\dagger} \upsilon =0$, $\sigma_z \upsilon=0$ where $\upsilon$
is an element of our algebra that is considered as algebra of
Hilbert-Schmidt operators in Fock space.

Let us remark that
an operator equation ${\bf K}{\bf L}=0$ is equivalent to condition  
that the
image of the operator ${\bf L}$ is contained in the kernel of ${\bf K}$. 
In other
words if we can solve the equation ${\bf K} {\bf v} =0$, where $\bf v$ 
is a vector
then we can also solve the operator equation ${\bf K}{\bf L}=0$, where $\bf L
$ is
an unknown operator.
This remark allows us to say that in the conditions above
we can consider $\upsilon$ as an element of Fock space;
if non-zero solution does not exist in this new setting it
does not exist in the old setting either. The elements of the Fock space
$\CH$ can be represented either as polynomials in $z_{0}, z_{1}$
(this representation we use below) or as $L^{2}$ functions on $\IR^{2}$
with coordinates $q_{0}, q_{1}$. 

 The equations for the vector $\upsilon$ can be written in holomorphic 
representation as follows:

\eqn\cndzmi{\matrix{
B_{0} \upsilon =   z_{0}  \upsilon &
B_{0}^{\dagger} \upsilon = {\zeta \over{4}} {\p \over{\p z_{0}}} \upsilon \cr
B_{1} \upsilon =  z_{1} \upsilon &
B_{1}^{\dagger} \upsilon = {\zeta \over{4}} {\p \over{\p z_{1}}} \upsilon  \cr
J \upsilon = 0 & I^{\dagger} \upsilon = 0\cr }}
The right column suggests that:
$$
\upsilon \sim 
e^{{{4z_{0} B_{0}}\over{\zeta}}  z_{0} B_{0}} \upsilon_{0} (z_{1})
$$
which is inconsistent with the left column equations.
We also say that the left column equations imply that the creation
operators have a finite-dimensional invariant subspace which is
impossible.
In the case $N=1$  the operator $\Delta_{z}$ 
is explicitly positive definite.

The question of existence of $\psi$ is addressed similarly:
write 
$\psi = \psi_{+} \oplus \psi_{-} \oplus \xi$, then \drdc\ assumes the form:
\eqn\drdcc{\eqalign{& (B_{0} + {\zeta \over{4}} {\p \over{\p \zb_{0}}}- z_{0}) 
\psi_{+} - (B_{1} + {\zeta \over{4}} {\p \over{\p \zb_{1}}} - z_{1}) 
\psi_{-} +I\xi  = 0 \cr
& (B_{1}^{\dagger} - {\zeta \over{4}} {\p \over{\p z_{1}}}
- \zb_{1})\psi_{+} +
 (B_{0}^{\dagger} - {\zeta \over{4}} {\p \over{\p z_{0}}}
- \zb_{0}) \psi_{-} + J^{\dagger}\xi = 0\cr}}
Here $\psi$ is a function which corresponds to  
an operator from our algebra.
The equation \drdcc\ 
can be rewritten in the form:
\eqn\drdcci{D_{A} \Psi = - \Xi}
for 
$$
\Psi = \pmatrix{ \psi_{+} \cr \psi_{-} \cr}, \quad 
\Xi = \pmatrix{- J^{\dagger}\xi \cr I \xi \cr}
$$
and $D_{A}$ being the Dirac operator in the gauge field $A_{\mu}dx^{\mu}$
where 
$$
\left( A_{\mu} - {\half} \omega_{\mu\nu} x^{\nu} \right)  dx^{\mu} = 
{4 \over{\zeta}} \left( -B_{1}^{\dagger} dz_{1}  
- B_{0}^{\dagger} dz_{0} 
+ B_{0}d\zb_{0}  + B_{1} d\zb_{1} \right) 
$$
The gauge field $A$ has constant curvature.

Now let us write down the equation \drdcc\ for vectors
in the Fock space, this time using $L^{2}$ represenation:
$\bf v = {\bf v}_{+}(q) \oplus {\bf v}_{-}(q) \oplus {\bf w} (q)$.

\eqn\drdccii{\eqalign{& (B_{0} + {\zeta \over{4}} {\p \over{\p q_{0}}}- q_{0}) 
{\bf v}_{+} - (B_{1} - {\zeta \over{4}} {\p \over{\p q_{1}}} + q_{1}) 
{\bf v}_{-} +I{\bf w}  = 0 \cr
& (B_{1}^{\dagger} - {\zeta \over{4}} {\p \over{\p q_{1}}}
- q_{1}){\bf v}_{+} +
 (B_{0}^{\dagger} - {\zeta \over{4}} {\p \over{\p q_{0}}}
- q_{0}) {\bf v}_{-} + J^{\dagger}{\bf w} = 0\cr}}

It has again the form \drdcci\ with $D_{A}$ being Dirac-like operator
in two dimensions. 
Let us assume that  the operator $D_A$ has no zero modes
(we hope that this
assumption can be justified in the framework of perturbation theory with
respect to $\zeta$). Then 
we can write a solution to \drdcci\ in terms
of the Green's function of the operator $D_{A}$. The space of
solutions to \drdccii\ is identified with the space of ${\bf w}$'s i.e.
the space of $W$-valued $L^{2}$ functions on $\IR^{2}$. 
Now, given the solution ${\bf v}_{+}({\bf w}) \oplus {\bf v}_{-}({\bf w})
 \oplus {\bf w}$
to \drdccii\ we construct a solution to 
\drdcci\ as follows: $\psi$ is supposed to map
$W \otimes \CH \otimes \CH \to (V \oplus V \oplus W) \otimes \CH \otimes \CH$
(where we used the fact that the algebra of Hilbert-Schmidt operators
can be identified with the Hilbert tensor product $\CH \otimes \CH$ and
that
$\CA_{\zeta} \subset \CH \otimes \CH$). Now, 
$$
\tilde \psi = {\bf w} \otimes {\bf w}^{\prime}  \mapsto
\left(  {\bf v}_{+}({\bf w}) \oplus {\bf v}_{-}({\bf w})
 \oplus {\bf w} \right) \otimes  {\bf w}^{\prime}  g
$$
does the job for any non-degenerate $g \in {\rm GL}_{k}(\CA)$. 
What remains is to normalize: $\psi = \left( 
\tilde \psi^{\dagger} \tilde \psi \right)^{-\half} \tilde \psi$.
This normalization
$\psi^{\dagger}\psi = 1$ reduces the freedom of the choice
of solutions to \drdccii\ 
to the unitary gauge transformations in $U(k)$. 

One can also rewrite the ASD equations for the gauge field on the 
non-commutative $\IR^{4}$ in 
terms of commutative functions  which are Wick symbols of the gauge fields. 
Let $A_{\mu}(x) $ be four  functions on $\IR^{4}$ and consider the equations:

\eqn\cmasd{F^{+}_{\mu\nu} =0}
with 
\eqn\nccrv{F_{\mu\nu, j}^{i} = 
\p_{\mu} A_{\nu, j}^{i} - \p_{\nu} A_{\mu, j}^{i} + A_{\mu, k}^{i} \star A_{\nu, j}^{k} - A_{\nu, k}^{i} \star A_{\mu, j}^{k}}

Thus, the non-commutative ASD equations can be thought of
the deformation of the ordinary ASD equations. 
The solutions to \cmasd\ are automatically the solutions to the deformed
Yang-Mills equations:
\eqn\dfrm{\p_{\mu} F_{\mu\nu} + A_{\mu} \star F_{\mu\nu} = 0}

\subsec{The completeness of the ADHM construction in the
 non-commutative case}

\lref\corrgodd{E.~Corrigan, 
P.~Goddard, ``Constructions of instanton monopole solutions and reciprocity'', Ann. Phys. (NY) {\bf 154} (1984), 253-279}

Just like in the case of ordinary $\IR^{4}$ one faces the question
- whether the full set of solutions of ASD equations 
on non-commutative $\IR^{4}$ is
enumerated by the solutions to the matrix equations \adhm. 
It is natural to try to imitate the arguments of \corrgodd\ and
express the matrices ${\bf B}_{\alpha}, I, J$ in terms of
solutions to the massless Dirac   equations in the instanton
background field. We simply sketch the relevant steps of the
construction without
giving any proofs. 

The setup is similar to the commutative case. Given a projective module $E$ over $\CA_{\zeta}$ one 
studies the associated the modules ${\bf S}_{\pm} = E \otimes_{\CA_{\zeta}} S_{\pm}$, 
$S_{\pm} = \IC^{2} \otimes \CA_{\zeta}$. 
Given a connection $\nabla$ in the module
$E$ we form the Dirac-Weyl  operators
\eqn\dwop{
D^{\dagger}: {\bf S_{+}} \to {\bf  S_{-}} , \quad
D: \bf  S_{-} \to \bf S_{+}}
by the standard formulae: $D^{\dagger} =  \nabla_{\alpha} \otimes \sigma^{\alpha}$, $D = \nabla_{\alpha} \otimes 
\bar\sigma^{\alpha}$
where $\sigma^{\alpha} : S_{+} \to S_{-}$, $\bar\sigma^{\mu}: S_{-} \to S_{+}$ are essentially the ordinary
Pauli matrices and $\sigma_{4} = \bar\sigma_{4} =1$.  One proves the identities:
\eqn\idepau{D D^{\dagger} = \Delta_{\nabla} \otimes {\bf 1} + F^{-}_{\mu\nu} \otimes \sigma^{\mu\nu}, \quad
D^{\dagger} D = \Delta_{\nabla} \otimes {\bf 1} + F^{+}_{\mu\nu} \otimes \bar\sigma^{\mu\nu}}
Here $\Delta_{\nabla}$ is the covariant Laplacian $E \to E$. 
Since there are no normalizable (see below) solutions to the equations
$\Delta_{\nabla} \phi =0$ then for the ASD $\nabla$ one concludes that there are no solutions to the equation $D \psi =0$. 
On the other hand, the index arguments predict, 
just like in the commutative case,
 the existence of $N$ normalizable (in the sense, described in the next paragraph)
zero modes of $D^{\dagger}$.  

Let $\psi^{i} \in E \otimes S_{+}$ be a solution to $D^{\dagger} \psi^{i} = 0$, $i=1, \ldots, N$. 
Then one may define a projection
(just like in the commutative case) of $x_{\alpha} \psi^{i}$ onto the space of zero modes of $\CD^{\dagger}$:
\eqn\prj{\left( x_{\alpha} \delta_{j}^{i}- {\bf B}_{\alpha,j}^{i} \right) \psi^{j} = D (\ldots)}
where ${\bf B}_{\alpha}$ is some matrix (with $\IC$-valued entries). 
We call $B_{0} = {\bf B}_{1} + i {\bf B}_{2}, B_{1} = {\bf B}_{3} + i {\bf B}_{4}$ etc. 

The matrices $I$, $J$ are recovered from the large $\vec x^2$ asymptotics of $\psi^{i}$. In order to explain what it
means in the non-commutative setting we represent the coordinates on the non-commutative $\IR^{4}$
as creation-annihilation 
operators acting in the auxilliary Fock space $\CH$. 
Then the operator $\psi^{i}$
has the corresponding Wick symbol $\tilde \psi^{i}$ which is simply a function on $\IR^{4}$ whose
large $\vec x^{2}$ limit 
is well-defined and is independent on the ordering prescription 
since for large $\vec x^{2} >> \zeta$ one may neglect the non-commutativity of
the coordinates. 

\newsec{Examples}

\subsec{Abelian instantons}

It is very interesting to study the case $k=1$. 
In the commutative case there are no solutions
to the abelian instanton equations except for the trivial ones. 
We shall see that for every $N$ there are non-trivial
abelian instantons in the non-commutative case. 

We need to solve the equations 
$\mu_{r}= \zeta \Id, \mu_{c} = 0$.
It can be shown that for $\zeta > 0$ the solution must have $J=0$ \nakheis.
Moreover, on a dense open set in $\CM_{\zeta}$ the matrices 
$B_{0}, B_{1}$ can be  diagonalized by a complex gauge
transformation\foot{We thank D.~Bernard for pointing out an error in the
earlier version of this paper}:
\eqn\dns{B_{\alpha} \to {\rm diag} 
\left( \beta_{\alpha}^{1}, \ldots, \beta_{\alpha}^{N} \right) .}
The eigenvalues $\beta_{\alpha}^{i}$ parameterize a set of $N$ points
on $\IR^{4}$. The actual moduli space is the resolution of singularities
of the symmetric product. Now suppose $(B_{0}, B_{1}, I)$ is a solution
to the ADHM equations. 
If we write the element of $V \oplus V \oplus W$ as $\psi= \psi_{0} \oplus 
\psi_{1} \oplus \xi$
then the equation $\CD^{\dagger} \psi=0$ is solved as:
\eqn\slt{\eqalign{\psi_{\alpha} & = - (B_{\alpha} - z_{\alpha})^{\dagger}
\delta^{-1} I \xi \cr 
\delta & = \sum_{\alpha = 0}^{1}
({B}_{\alpha} - {z}_{\alpha} )({B}_{\alpha}^{\dagger} -
 {\zb}_{\alpha} )  \cr}}
The operator $\xi$ is determined from the equation $\psi^{\dagger}\psi =1$:
\eqn\xe{\xi = \left( 1 + I^{\dagger} \delta^{-1} I \right)^{-\half}}
The connection $A = \psi^{\dagger} d\psi$ has  Yang's form:
\eqn\yng{A = \xi^{-1} \pb \xi - \p \xi \xi^{-1}}
where $\p = dz_{\alpha} {\p \over{\p z_{\alpha}}}$, 
$\pb = d\zb_{\bar\alpha} {\pb \over{\p \zb_{\bar\alpha}}}$.
Explicitly:
$$
A = \xi^{-1} d \xi + \xi^{-1} \alpha \xi,
$$
where the ``gauge transformed'' connection $\alpha$ is equal to:
\eqn\ccnt{\eqalign{\alpha & = \xi^{2} \p \xi^{-2} = \cr
 {1\over{ 1 + I^{\dagger} \delta^{-1} I}}  &
I^{\dagger} \delta^{-1} ( B_{\alpha}^{\dagger} - {\zb}_{\alpha} ) 
dz_{\alpha} \delta^{-1} I \cr}}
For $N=1$: $\xi = \left( {{d - \zeta/2}\over{d + \zeta/2}}
\right)^{\half}$, 
$d = z_{0}\zb_{0} + \zb_{1} z_{1}$,
and the gauge field $\alpha$ is explictly
non-singular if the correct ordering is used:
\eqn\al{\alpha = {1\over{d (d+\zeta/2)}} 
\left( \zb_{0} dz_{0} + \zb_{1} dz_{1}\right)}
One can also compute the curvature:
\eqn\crvt{\eqalign{
& F_{A} = {{\zeta}\over{(d- \zeta/2) d (d + \zeta/2)}} \left( f_{3} \left( dz_{0}d\zb_{0} - dz_{1}d\zb_{1} \right)+
f_{+} d\zb_{0}dz_{1} + f_{-} d\zb_{1} dz_{0} \right) \cr
&  f_{3} =  z_{0}\zb_{0} - z_{1}\zb_{1}, \quad f_{+} = 2 z_{0}\zb_{1}, \quad
f_{-} = 2 z_{1} \zb_{0}\cr}}
The factor ${{\zeta}\over{(d- \zeta/2) d (d + \zeta/2)}}$ has a singularity
at the state $\vert 0,0\rangle$ but it is projected out
since $f_{3,\pm}$ always have $\zb_{0}$ or $\zb_{1}$ on the right.
The action density is given by
\eqn\actndsn{{\hat S} = -{1\over{8\pi^{2}}} F_{A}F_{A} = {{\zeta^{2}}\over{4\pi^{2}}}
{1\over{d^{2} (d - \zeta/2) (d +\zeta/2)}} \Pi}
where $\Pi = 1 - \vert 0,0 \rangle \langle 0,0 \vert$. 
We may define the total action as
$$
(\zeta \pi)^{2}  {\Tr}_{\CH} {\hat S} = 
4 \sum_{N=1}^{\infty} {1\over{N(N+1)(N+2)}}  = 1
$$

\subsec{'t Hooft solutions}
\lref\raja{R.~Rajaraman, ``Solitons and Instantons'', North-Holland, 1987}
It is also relatively easy to describe the non-commutative analogues of 't Hooft solutions for $k=2$:
\eqn\ans{
A_{\mu} = i \Sigma^{\mu\nu} {\Phi}^{-1} \p_{\nu} \Phi
}
where $\Sigma^{\mu\nu}$ is self-dual in $\mu\nu$ and takes values in
traceless two by two Hermitian matrices.
In the commutative case the ASD conditions boil down to the 
Laplace equation on $\Phi$ \raja. 
In the non-commutative case the potential trouble comes from the
term in the curvature $F_{\mu\nu}$:
$$
\{ \Sigma^{\mu\alpha}, \Sigma^{\nu\beta} \} [ J_{\alpha}, J_{\beta}]
$$
with $J_{\alpha} = \Phi^{-1} \p_{\alpha} \Phi$. It is easy to show that
the problematic piece is equal to 
$$
[J_{\mu}, J_{\nu}] - 
{\half}\epsilon_{\mu\nu\alpha\beta} [ J^{\alpha}, J^{\beta}]
$$
which is explicitly anti-self-dual. Hence we have shown  that the ansatz
\ans\ works in the non-commutative case if $\Phi$ obeys the Laplace
equation
which is now to be solved
in the non-commutative setting. 
A solution looks exactly like  the commutative
ansatz:
\eqn\tha{\Phi = 1 +\sum_{i=1}^{N} 
{{\rho_i^2}\over{\vert \vec x - \vec \beta_{i} \vert^{2}}}}
where now the components of $\vec x = (x^{\mu})$
 represent the non-commuting coordinates. 

One might wonder what are the properties of $\Phi$ viewed as an operator in a 
Fock space, where
$x_{\alpha}$'s are realized as the creation and annihilation operators.
By acting on a vacuum state $\vert 0,0\rangle$ in the occupation number representation 
$\Phi$ creates a sum of the coherent states. We may represent
$\Phi$ as follows:
$$
\Phi = 1 + \sum_{i=1}^{N} \rho_{i}^{2} \upsilon_{i}
$$
where $\upsilon_{i} = e^{-\vec x^{\vee} \cdot \vec \beta_i} 
\Upsilon e^{ \vec x^{\vee} \cdot \vec \beta_i}$, 
$\Upsilon = {1\over{\vec x^{2}}}$. 
 Let us now prove that 
each $\upsilon_{i}$ obeys Laplace equation. Since the authomorphismes in $\lieg = \IR^{4}$  are internal and
they commute it is sufficient to prove that $\Upsilon$ obeys the Laplace equation. 
For the latter it is enough to check that
$$
\vec x^{2} \Upsilon + \Upsilon \vec x^{2} = 2 \vec x \Upsilon \vec x
$$
which is true for $\Upsilon = {1\over{\vec x^2}}$. The operator $\Upsilon$ is diagonal in the occupation number
representation and its eigenvalue on a state $\vert n_{0}, n_{1} \rangle$ is equal to
${1\over{{\zeta \over{2}} \left( n_{0} + n_{1} + 1 \right)}}$.

\newsec{Relation to Matrix description of six dimensional $(2,0)$ theory}

\lref\sxa{O.~Aharony, M.~Berkooz, S.~Kachru, N.~Seiberg, E.~Silverstein,
`` Matrix Description of Interacting Theories in Six Dimensions'', 
hep-th/9707079}

In the recent papers \abs\sxa\ the proposal for Matrix description of
six dimensional $(2,0)$ superconformal theory of
$k$ coincident fivebranes has been made. The theory which arises on
the worldvolume of $k$ coincident fivebranes has the property that when it
is compactified on a circle it becomes $U(k)$ gauge theory in $4+1$ dimension
which the coupling which goes to zero as the radius goes to zero.
So, by compactifying the theory on a light-like
circle one gets the gauge theory with zero coupling. Of course, in a given
instanton sector the only surviving
gauge configurations are instantons.  The theory becomes a quantum mechanics
on the moduli space of instantons. But, the latter space has singularities coming from
the point-like instantons and at these singularities a second branch of the theory,
corresponding to the emmission or absorption of $D0$-branes develops.
The theory becomes interacting with the bulk degrees of freedom which makes
it harder to study. The proposal of \abs\ was to deform the instanton moduli
space by turning on a $B$ field along
four dimensions which serves as a FI term $\zeta$ in \nmdl. The interpretation of our paper is that
turning on the $B$ field effectively makes $\IR^{4}$ non-commutative 
and the instantons
live on it.

Now, the problem of instantons shrinking to zero size is cured by quantum fluctuations! Indeed,
the position of the point-like  instanton is smeared over a region of size $\sim \zeta$ which
makes it no longer point-like. So the Higgs branch becomes smooth
hyperkahler manifold and the theory becomes six dimensional (at least decoupled
from the bulk degrees of freedom).

\newsec{Future directions}

In this section we briefly sketch a few directions of future research.

\subsec{Nahm's transform}
\lref\vbaal{P.~ van Baal, P.~ Braam, ``Nahm's transformation for instantons'' , Comm.
Math. Phys. {\bf B} 122 (1989) 267-280}
Let us define an $n$-dimensional noncommutative torus ${\cal A}_{\theta}$
as an associative involutive algebra with unit generated by the unitary 
generators $U_1,...,U_n$ obeying 
$$
U_{\mu}U_{\nu}=e^{i\theta_{\mu \nu}}U_{\nu}U_{\mu}
$$
Here $\theta_{\mu \nu}$ is an antisymmetric tensor; we can also consider
it as a $2$-form on ${\bf R}^n$. The infinitesimal authomorphismes
 $\delta_{\alpha}
U_{\beta} = \delta_{\alpha \beta} U_{\beta}$ generate the abelian
 Lie algebra $L_{\theta}$ acting  on ${\CA}_{\theta}$. As before
we use
$L_{\theta}$ to define the notion of connection in a 
${\CA}_{\theta}$-module. 
We always consider modules equipped with Hermitian
inner product and Hermitian connections. 
\lref\prep{A.~Astashkevich, N.~Nekrasov, A.~Schwarz, in preparation}
We will construct a
generalization of Nahm's transform \vbaal\ relating connections on 
${\CA}_{\theta}$-modules to connections on 
${\CA}_{\hat\theta}$-modules \prep. Here ${\CA}_{\theta}$ and 
${\CA}_{\hat\theta}$ are two four-dimensional noncommutative tori.
To define a non-commutative generalization of Nahm's transform we need a
$({\CA}_{\theta}, {\CA}_{\hat\theta})$-module ${\CP}$ with
${\CA}_{\theta}$ -connection $\nabla_{\alpha}$ and 
$\CA_{\hat\theta}$-connection $\hat{\nabla}_{\mu}$. 
The fact that  $\CP$ is   $({\CA}_{\theta},{\CA}_{\hat\theta})$-module 
means that ${\CP}$ is a left ${\CA}_{\theta}$-module and a right 
${\CA}_{\hat{\theta}}$-module; we assume that $(ax)b=a(xb)$ for 
$a\in {\CA}_{\theta},\  b\in {\CA}_{\hat {\theta}},\  x\in {\cal P}$. 
In other words, ${\CP}$ can be considered as 
${\cal A}_{\theta\oplus \hat {\theta}}$-module, 
where ${\cal A}_{\theta \oplus \hat {\theta}}$ is an
eight-dimensional noncommutative torus corresponding to the $2$-form
$\theta \oplus \hat {\theta}$ on ${\bf R}^8$. Assume that the
commutators $[\nabla _{\alpha}, \nabla _{\beta}],\  [\hat {\nabla}_{\mu},
\hat {\nabla}_{\nu}],\  [\nabla _{\alpha},\hat {\nabla}_{\mu}]$ are
$c$-numbers: 
\eqn\cmmt{[\nabla _{\alpha}, \nabla _{\beta}]=\omega _{\alpha \beta},\  [\hat
{\nabla}_{\mu}, \hat {\nabla}_{\nu}]=\hat {\omega}_{\mu \nu},\  [\nabla
_{\alpha},\hat {\nabla}_{\mu}]=\sigma _{\alpha \mu}.}
The condition \cmmt\ 
implies that the curvature of connection $\nabla _{\alpha} \oplus \hat
{\nabla}_{\mu}$ on $\CA_{\theta \oplus \hat\theta}$ is constant.
One more assumption is that $\nabla_{\alpha}$ commutes with
multiplication by elements of ${\CA}_{\hat\theta}$ and 
$\hat\nabla_{\alpha}$ commutes with multiplication by elements of 
${\CA}_{\theta}$. Of course, the module ${\CP}$ and the connections  
$\nabla_{\alpha},  \hat {\nabla}_{\mu}$ 
obeying the above requirements exist
only under certain conditions on $\theta ,\hat\theta$. 
For 
any right ${\CA}_{\theta}$-module $R$ with connection 
$\nabla_{\alpha}^R$ we consider Dirac operator $\CD =\Gamma^{\alpha}
(\nabla_{\alpha}^R + \nabla_{\alpha})$ 
acting in the tensor product 
$$(R\otimes_{{\CA}_{\theta}}{\CP})\otimes S.$$
(To define $\Gamma$-matrices we introduce inner product in $L_{\theta}$
and in $L_{\hat {\theta}}$.) This operator commutes with multiplication by
the 
elements of ${\CA}_{\hat\theta}$ hence the space of zero modes
of $\CD$ 
can be regarded as ${\CA}_{\hat\theta}$-module; we denote it
as $\hat{R}$. The connection $\hat\nabla_{\mu}$ induces a
connection $\hat\nabla_{\mu}^{\prime}$ on 
$$(R\otimes_{{\CA}_{\theta}}{\CP})\otimes S.$$
We define a connection $\nabla_{\mu}^{\hat{R}}$ on $\hat{R}$ as 
$P\hat\nabla_{\mu}^{\prime}$ where $P$ is the orthogonal projection:
 $$(R\otimes_{{\CA}_{\theta}}{\CP})\otimes S \to \hat{R}$$

The above construction can be regarded as generalized Nahm's transform. To
prove that its properties are similar to the properties of standard Nahm's
transform we should impose additional conditions on module ${\CP}$ and
connections $\nabla_{\alpha},\  \hat\nabla_{\mu}$. It is 
sufficient to assume that  
$\sigma _{\alpha \mu}$ determines a non-degenerate pairing between
$L_{\theta}$ and 
 $L_{\hat {\theta}}$.  
Then we can use this pairing to define an inner
product in 
 $L_{\hat {\theta}}$.
A connection $\nabla_{\alpha}^R$ is an analogue of ASD
connection if its curvature $F_{\alpha \beta}$ obeys $F_{\alpha
\beta}^{+}+ \omega_{\alpha \beta}\cdot 1=0$.  
(This condition is equivalent
to antiselfduality of the connection $\nabla_{\alpha}^R+\nabla_{\alpha}$.) 
Then one can prove that the curvature $\hat{F}_{\mu \nu}$
satisfies $\hat{F}_{\mu \nu}^{+}- \hat\omega_{\mu \nu } \cdot 1=0$.

The forms $\omega, \hat\omega$ obey certain quantization conditions
which depend on $\theta, \hat\theta$. We plan to return to this
issue elsewhere.

Thus, Nahm's transform maps the modified instantons on one non-commutative 
torus to the modified
instantons on the other non-commutative torus. It can be thought of
an analogue of 
 Morita equivalence.
The last notion is usually used in the context of equivalence of algebras 
and the categories of left- and/or
right- modules over them.\foot{
Two algebras $A, B$ are called Morita equivalent if there exist the $(A,B)$ module $E$ and $(B, A)$ module
$F$ s.t 
$E \otimes_{B} F \approx A, \quad F \otimes_{A} E \approx B$
For example, every algebra $A$ is Morita equivalent to the 
algebra ${\rm Mat}_{N}(A)$ of matrices with 
coefficients in $A$. 
Given two Morita equivalent algebras one may construct for any 
$A$-module $\CE$ a $B$-module
$\CF$ and vice versa: 
$\CF = F \otimes_{A} \CE$ .

The fact that  the holomorphic counterpart  of Nahm's transform -
the Fourier-Mukai transform -  can be understood  in the framework
of generalized Morita
equivalence has been pointed
out to us by M.~Kontsevich.}. 
In the context of gauge theories we study the modules with connections. 
 Nahm's construction  gives correspondence between such modules;
 modules with ASD connections
are mapped to each other.

The instantons on the non-commutative torus appear in the problem of
$D0$-branes bound to $D4$-branes wrapping the four-torus $A = T^{4}$
with the $B$-field
\lref\mdsn{M.~Douglas, G.~Moore, N.~Nekrasov, S.~Shatashvili, unpublished}
turned on.  $T$-duality maps the torus $A$ to its dual and 
the natural conjecture \mdsn\  is that on the level
of low-energy fields it maps the instantons on $A$ to those on $A^{\vee}$. 
There exist stronger conjectures about the moduli space
of instantons on the four-torus which fulfill the base of the
constructions fo six dimensional interacting string theories
\lstrings\ and also provide a heuristic test of $U$-dualities \dvafa. 
One assumes that $\CM_{N, U(k)} = {\rm Sym}^{kN} T^{4}$ and studies the 
two dimensional sigma model  with target $\CM_{N,U(k)}$. 
Since the symmetric product is
 an orbifold of the hyperkahler space one may perturb
the theory by the 
marginal operators 
responsible for blowing up the singularities of the orbifold. 
As was argued in
\witthig\ such a perturbation breaks the natural symmetries
of the problem (and cannot be interpreted as responsible for
interactions of little strings) . 
We have argued that given a non-commutative torus 
(or non-commutative
$\IR^{4}$) where the relevant $SU(2)$ symmetry is already broken by
the non-commutative deformation the moduli space of instantons 
is already smooth and  
the decoupling arguments can be applied.

\subsec{Instantons on non-commutative ALE spaces}
\lref\mckay{J.~McKay, ``Graphs, Singularities and Finite groups'', Proc. 
Sympos. Pure Math. {\bf 37} AMS (1980), 183-186}
The asymptotically locally eucldean (ALE) manifolds
$X_{\Gamma}$ are the hyperkahler resolutions of singularities
of the orbifold $\IC^{2}/{\Gamma}$ for $\Gamma \subset SU(2)$ being a
discrete subgroup. The subgroups correspond to $ADE$ Lie groups $G$
via McKay's correspondence\mckay. The space $X_{\Gamma}$ depends on 
$r = {\rm rk}G$ parameters $\vec\zeta_{i}\in \IR^{3}, i=0, \ldots, r$,
$$
\sum_{i} \vec \zeta_{i} = 0
$$
\lref\quiv{M.~Douglas, G.~Moore, ``D-branes, Quivers and
ALE Instantons,'' hep-th/9603167.}%
The spaces $X_{\Gamma}$ as well as the moduli spaces
of $U(N)$ instantons on $X_{\Gamma}$ can be constructed
via hyperk\"ahler reductions of vector spaces \kronheimer\KN.
These constructions were interpreted in  \quiv\ as originating
from the gauge theory on $v$ $Dp$-branes, put at the orbifold
point of $\IC^{2}/{\Gamma}$ inside the $w$ $D (p+4)$-branes.
It has been noticed in \quiv\ that in principle the condition
$\sum_{i} \vec\zeta_{i} = 0$ can be relaxed by turning on a self-dual
part of $B_{\mu\nu}$ along $\IC^{2}/{\Gamma}$. 
As before, we interpret this as the process
of  going to the non-commutative 
ALE space with instantons on it. The ADHM construction of \KN\ generalizes
straightforwardly to this case provided the original vector space
$V = {\rm Hom}(\IC^{\Gamma}, \IC^{\Gamma} \otimes \IC^{2})^{\Gamma}$ 
whose reduction yields the ALE space is replaced by its non-commutative
deformation:
$$
V^{\rm q} = {\rm Hom}( \IC^{\Gamma} , \IC^{\Gamma} \otimes 
\CA_{\zeta})^{\Gamma}
$$
\lref\csev{L.~Baulieu, A.~Losev, N.~Nekrasov, hep-th/9707174}

\subsec{Instantons and holomorphic bundles on non-commutative surfaces}
\lref\gieseker{For Gieseker's compactification, see
C~Okonek, M.~Schneider, H.~Spindfler, ``Vector bundles on complex
 projective spaces'', Progress in Math., Birkhauser, 1980}

 It is well-known that solutions to the instanton equations 
on the complex surface
determine the holomorphic structure in the bundle where the gauge field is defined.
In fact, Donaldson-Uhlenbeck-Yau theorem establishes an equivalence
between the moduli space of stable in a certain sense holomorphic bundles
over a surface $S$ 
and solutions to the instanton equations on $S$ \dnld\DoKro. 
The holomorphic bundle $\CE$ defines a sheaf of its sections, 
and can be replaced by this
sheaf for many purposes. In fact, not every sheaf $\CF$ comes from a bundle - 
for this it must be what is called a locally-free sheaf (the term free means that the  sections 
of $\CF$ 
form a free module over the sheaf $\CO$ of holomorphic functions on $S$). 
By relaxing the condition of being locally free
but insisting on being the torsion free one gets a generalization
of the notion of a holomorphic bundle.  However, it was not clear how
to obtain corresponding generalization of the notion of instanton. 
Of course, the consideration of torsion free sheaves allows us to 
compactify the {\it moduli space} of instantons \gieseker\nakheis. 
But it is by no means clear whether the compactification
can be achieved within a gauge theory. It has been conjectured in \vw\
that such a compactification occurs in string theory.

We claim that {\it there exists a generalization of the instanton 
field} for the torsion free sheaves. 
This is simply the gauge field on the {\it non-commutative surface} $S$.

The importance of the spaces $\CM_{\zeta}$ has been appreciated in the 
context of string duality 
and field theory a while ago (see \avatar\hm\dvafa\abs). We hope that the 
proposed interpretation will help
to find further applications as well as justify various assertions needed 
for establishing string/field dualities.

In our presentation the non-commutativity of $\IR^{4}$ entered only
through the construction of 
the connection in a given bundle. 
The holomorphic bundle
underlying the instanton could be described in purely commutative
terms. This relies on the fact that with our choice of the Poisson
structure the subalgebra $\CO$ of holomorphic functions on
$\IC^{2}$ is commutative. It is interesting to study the picture in other
complex structures. In fact, slightly redundant but interesting
deformation of ADHM equations is the 
three-parametric one:
\eqn\adhmdf{\mu_{r} = \zeta_{r} \Id, \quad \mu_{c} = \zeta_{c} \Id}
where $\zeta_{r} \in \IR, \zeta_{c} \in \IC$. It is customary
to combine $(\zeta_{r}, \zeta_{c}, \bar\zeta_{c})$ into a three
vector $\vec\zeta \in \IR^{3}$. 

The specifics of $\IR^{4}$ is that one can always get rid of $\zeta_{c}$
by 
appropriate rotation in the $SU(2)_{R}$ global group.
Let us not do it but rather look at the complex equation
$\mu_{c} = \zeta_{c} \Id$:
\eqn\adhmmdf{\tau_{z} \sigma_{z} = 0, {\quad} {\rm iff} \quad
[z_{0}, z_{1} ] = - \half\zeta_{c}}
Thus the $\zeta_{c}$-deformation allows one to construct a 
coherent sheaf over the non-commutative $\IC^{2}$ as the cohomology
of the complex of sheaves:
\eqn\cmplx{\matrix{ & & & \sigma_{z} & & \tau_{z} & & &  \cr
0 & \to & V \otimes \CO_{\zeta_{c}}  & \to & \left( V \otimes \IC^{2} \oplus W \right) 
\otimes \CO_{\zeta_{c}} & \to & V \otimes \CO_{\zeta_{c}} & \to & 0,\cr}}
where 
$\CO_{\zeta_{c}}$ is the associative algebra  generated by
$z_{0}, z_{1}$ obeying the relation
$[z_{0}, z_{1}] = - {\half}\zeta_{c}$ and $\tau_{z}, \sigma_{z}$ are given by the same formulae \de.

\newsec{Acknowledgements}

We are grateful 
to the organizers of the Workshop on ``Dualities in String Theory'' 
 at ITP at
Santa Barbara and 
especially 
to M.~Douglas, D.~Gross, W.~Lerche and H.~Ooguri
for providing a very stimulating athmosphere. 
We are 
 grateful to A.~Astashkevich, T.~Banks, D.~Bernard, 
V.~Fock, A.~ Gerasimov, M.~Kontsevich, 
A.~Losev, G.~Moore , M.~Rieffel,
A.~Rosly, K.~Selivanov, S.~Sethi and  S.~Shatashvili 
for useful discussions.
The research of N.~Nekrasov was supported by Harvard Society of 
Fellows,
partially by NSF under  grant
PHY-92-18167, partially by RFFI under grant 96-02-18046 and partially
by grant 96-15-96455 for scientific schools. 
The research of 
A.~Schwarz
was partially supported by NSF under grant DMS-9500704.

\lref\berkooz{M.~Berkooz, to appear}

As our paper was ready for publication we have learned about the paper
\berkooz\ which will adress
certain issues concerning 
gauge theories on  non-commutative $T^{4}$.

\listrefs
\bye